\documentclass{article}
\usepackage{spconf,amsmath,graphicx,hyperref}
\usepackage{cite}
\usepackage{,amssymb,amsfonts}
\usepackage{algorithmic}
\usepackage{graphicx}
\usepackage{textcomp}
\usepackage{multirow}
\usepackage{booktabs, threeparttable}
\usepackage{xcolor}

\title{Discrete Diffusion for Generative Modeling of Text-Aligned Speech Tokens}
\name{Pin-Jui Ku$^{1,2}$,
      He Huang$^{1}$,
      Jean-Marie Lemercier$^{1}$,
      Subham Sekhar Sahoo$^{1,3}$,
      Zhehuai Chen$^{1}$,
      Ante Juki\'{c}$^{1}$}
\address{$^{1}$NVIDIA, USA \;
$^{2}$Georgia Institute of Technology, USA\;
$^{3}$Cornell Tech, USA
}

\begin{document}
\ninept
\maketitle
\begin{abstract}
This paper introduces a discrete diffusion model (DDM) framework for text-aligned speech tokenization and reconstruction. By replacing the auto-regressive speech decoder with a discrete diffusion counterpart, our model achieves significantly better reconstruction quality, stronger ASR performance, and faster inference. We provide a comprehensive analysis of applying DDMs to speech reconstruction, examining sampler choices, inference steps, and robustness to length-scale estimation errors. Furthermore, we improve the original TASTE by systematically comparing vector quantization modules, showing that FSQ yields up to a 35\% relative WER reduction and +0.14 UT-MOS improvement over RVQ for AR models, while also enhancing DDM performance. Our model generates speech in just 10 denoising steps and even supports single-step generation with only minor quality degradation.
\end{abstract}
\begin{keywords}
Discrete Diffusion Model, Speech Tokenization, Speech Reconstruction
\end{keywords}
\section{Introduction}
\label{sec:intro}

In recent years, \emph{discrete diffusion models} (DDMs) have emerged as a powerful generative modeling framework for language and sequence modeling. Originating from the Discrete Denoising Diffusion Probabilistic Models (D3PM)~\cite{austin2021structured}, this line of research has advanced rapidly through extensions such as the Masked Diffusion Language Model (MDLM)~\cite{sahoo2024simple}, Discrete Flow Matching~\cite{gat2024discrete}, ReMDM~\cite{wang2025remasking}, and Edit-Flow~\cite{havasi2025editflow}. Unlike traditional autoregressive (AR) models that require strictly sequential decoding, DDMs perform inference through iterative but inherently parallelizable denoising steps. This property enables efficient sampling, flexible trade-offs between speed and quality, and greater controllability in generation. Such advantages have driven their adoption in large-scale language models including LLaDA~\cite{nie2025llada} and Mercury~\cite{labs2025mercury}.

While these advances have primarily focused on language modeling, the application of DDMs to \emph{speech-related} tasks remains underexplored. Recent works such as Diffusound~\cite{Yang2023Diffsound} demonstrated the potential of replacing AR decoders with discrete diffusion for text-to-sound generation, operating directly on quantized mel-spectrogram tokens. More recent efforts, including NaturalSpeech3~\cite{ju2024naturalspeech3} for text-to-speech (TTS), automatic speech recognition (ASR)\cite{kwon2025whisfusion}, token-based audio inpainting\cite{dror2025token}, and bandwidth extension~\cite{Fang2025vector}, further highlight the versatility of DDMs for speech. However, these studies do not leverage the latest advances in training and sampler design~\cite{wang2025remasking,kim2025train,yu2025dimple,ben2025accelerated}, which greatly improve efficiency, scalability, and controllability.

In this work, we present the first comprehensive study of DDMs for speech tokenization and reconstruction. Our approach builds on the TASTE framework~\cite{tseng2025taste}, which aligns speech tokens with transcription tokens to produce highly efficient, text-aligned, low-bitrate speech representations well-suited for spoken language models. However, the original TASTE relies on an AR speech decoder to predict S3 token sequences~\cite{du2024cosyvoice1} before time-domain signal reconstruction, resulting in inefficiency and suboptimal generation. To address this limitation, we propose to replace the AR decoder with a DDM-based decoder and conduct a thorough analysis of its effectiveness on speech reconstruction tasks.

In summary, this work makes three main contributions. First, we present the first application of DDMs to speech tokenization and reconstruction, achieving higher reconstruction quality and significantly faster inference than the AR-based TASTE baseline. Second, we enhance TASTE’s performance through a systematic analysis of quantizer design. Finally, we provide a comprehensive evaluation of inference settings—including sampler choice, number of steps, and robustness to length estimation errors—offering insights and guidance for deploying DDMs in speech applications.

\section{Approach}
\label{sec:models}
\subsection{Discrete Diffusion Models}
\label{subsec:ddm-intro}

The foundation of discrete diffusion models can be traced back to D3PM introduced by Austin \emph{et al.}~\cite{austin2021structured}, which extended the continuous diffusion framework to categorical state spaces and provided a principled way of adapting diffusion models to discrete domains. A widely studied special case of D3PM is the \emph{masking} or \emph{absorbing state} discrete diffusion process. Its forward process $q$ can be formally defined as follows. Let $\mathbf{x} \in \{0,1\}^{|V|} \subset \Delta^{|V|}$ denote a one-hot vector representing a clean token, where $|V|$ is the vocabulary size and $\Delta^{|V|}$ is the probability simplex over this vocabulary. Given a time step $t \in [0,1]$, the marginal distribution of an intermediate latent variable $\mathbf{z}_t \in \{0,1\}^{|V|}$ is
\begin{equation}
    q(\mathbf{z}_t \mid \mathbf{x}) = \mathsf{Cat}\big(\mathbf{z}_t; \alpha_t \mathbf{x} + (1 - \alpha_t)\mathbf{m}\big),
    \label{eq:forward}
\end{equation}
where $\mathsf{Cat}\left(. ; \mathbf{p} \right)$ is the categorical distribution over $|V|$ classes with probabilities $\mathbf{p} \in \Delta^{|V|}$, $\mathbf{m}$ is a one-hot vector corresponding to a special $[\text{MASK}]$ token, and $\alpha_t \in [0,1]$ is a strictly decreasing schedule with $\alpha_0 \approx 1$ and $\alpha_1 \approx 0$. For sequences of $L$ tokens, we denote clean and noisy sequences as $\mathbf{x}^{(1:L)}$ and $\mathbf{z}_t^{(1:L)}$, with individual tokens represented as $\mathbf{x}^{(\ell)}$ and $\mathbf{z}_t^{(\ell)}$ for $\ell \in \{1, \ldots, L\}$.

Building on this formulation, Sahoo \emph{et al.}~\cite{sahoo2024simple} proposed the MDLM as an improved version of the D3PM absorbing-state process. By introducing zero masking probabilities and carry-over properties for unmasked tokens, MDLM simplifies the reverse posterior \mbox{$q(\mathbf{z}_s \mid \mathbf{z}_t, \mathbf{x})$} into
\begin{equation}
    q(\mathbf{z}_s \mid \mathbf{z}_t, \mathbf{x}) 
    = \mathsf{Cat}\!\left(\mathbf{z}_s; 
        \frac{\alpha_s - \alpha_t}{1 - \alpha_t}\mathbf{x} 
        + \frac{1 - \alpha_s}{1 - \alpha_t}\mathbf{z}_t
    \right),
    \label{eq:posterior}
\end{equation}
where $0 < s < t$ denotes a preceding time step.  In this setting, a denoising neural network $\mathbf{x}_\theta$ parameterized by $\theta$ is used to implement the reverse unmasking process:
\begin{align*}
    p_\theta(\mathbf{z}_s|\mathbf{z}_t) 
    &= q(\mathbf{z}_s \mid \mathbf{z}_t, \mathbf{x}_\theta(\mathbf{z}_t)) \\
    &= \begin{cases}
        \mathsf{Cat}(\mathbf{z}_s; \mathbf{z}_t), & \mathbf{z}_t \neq \mathbf{m}, \\[6pt]
        \mathsf{Cat}\!\left(\mathbf{z}_s; 
            \dfrac{(1-\alpha_s)\mathbf{m} + (\alpha_s - \alpha_t)\mathbf{x}_\theta(\mathbf{z}_t)}{1 - \alpha_t}
        \right), \text{\hspace{-3mm}} & \mathbf{z}_t = \mathbf{m}.
    \end{cases}
\end{align*}
The model is trained using the following objective:
\begin{equation}
    \mathcal{L}
    = \mathbb{E}_{q} \int_{t=0}^{1} 
    \frac{\alpha_t'}{1 - \alpha_t} 
    \sum_{\ell} \log \big\langle 
    \mathbf{x}_\theta^{(\ell)}(\mathbf{z}_t^{1:L}, t), \mathbf{x}^{(\ell)} 
    \big\rangle \, \text{d}t.
    \label{eq:ddm-objective}
\end{equation}


\subsection{TASTE: Text-Aligned Speech Tokenization}
\label{subsec:taste}

TASTE~\cite{tseng2025taste} is a recently proposed framework for \emph{text-aligned speech tokenization}, designed to enable more effective joint speech and text modeling for spoken language model applications. Its key idea is to align speech tokens with their corresponding text tokens during tokenization, thereby creating a one-to-one correspondence between text and speech representations. This framework addresses the common \emph{token length mismatch problem} between speech and text modalities~\cite{defossez2024moshi, fang2024llama, xie2024mini}, which has been a major challenge for large spoken language models.

\begin{figure}
    \centering
    \includegraphics[width=\linewidth]{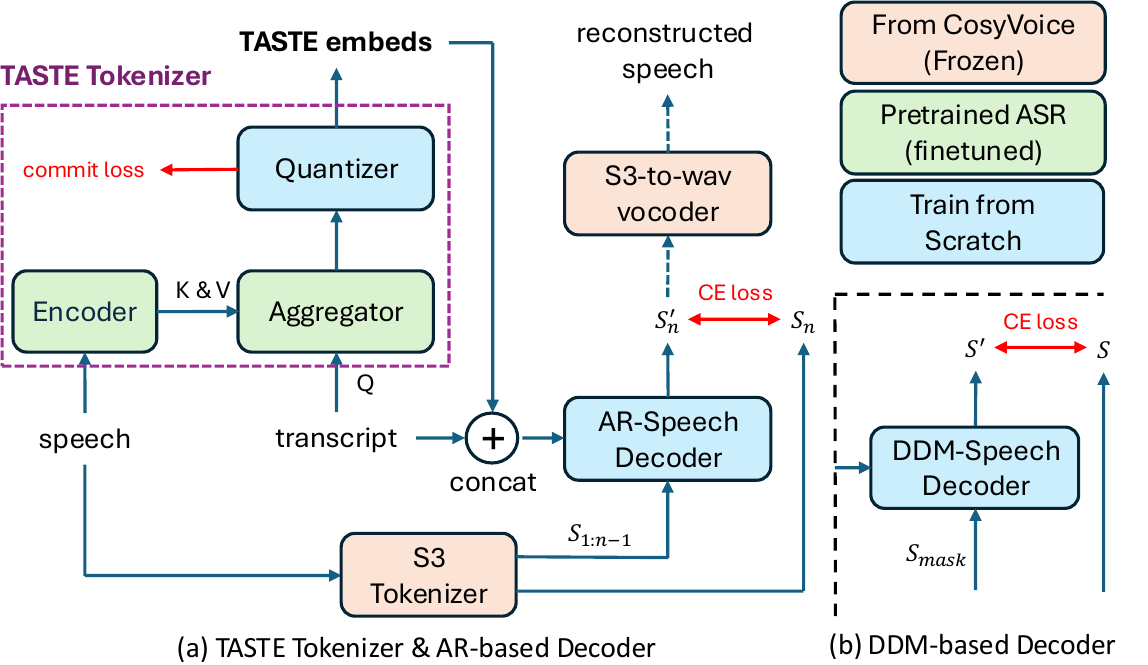}
    \vspace{-6mm}
    \caption{Overview of (a) the original TASTE framework with an AR speech decoder and (b) the proposed DDM-based decoder.}
    \vspace{-4mm}
    \label{fig:TASTE-illustration}
\end{figure}

As illustrated in Fig.~\ref{fig:TASTE-illustration}(a), TASTE consists of two main components: a \textbf{TASTE tokenizer} which produces text-aligned speech tokens and an AR \textbf{speech decoder} which reconstructs speech from the transcript and the aligned TASTE tokens.
The tokenizer consists of an encoder, an aggregator, and a quantizer. The encoder, initialized from the Whisper encoder~\cite{radford2022whisper}, extracts acoustic features that are processed by an \emph{aggregator}, initialized from the Whisper decoder, to align the acoustic features with the corresponding text tokens through cross-attention. Finally, a quantizer based on residual vector quantization (RVQ)~\cite{zeghidour2021soundstream} discretizes the aligned acoustic representations into the TASTE speech tokens.  

For reconstruction, TASTE employs a Transformer-based speech decoder that conditions on both the transcript and the continuous TASTE embedding vectors (rather than their discrete indices). The decoder autoregressively predicts S3 tokens from CosyVoice~\cite{du2024cosyvoice1}, which serve as intermediate representations between TASTE tokens and waveforms. These predicted S3 tokens are then converted into mel-spectrograms using a pretrained flow-matching S3-to-mel converter, followed by waveform generation with a HiFi-GAN vocoder~\cite{Kong2020hifigan}. The overall training objective combines two losses: (1) a cross-entropy loss between the predicted and ground-truth S3 tokens, and (2) a commitment loss from the RVQ quantizer to encourage stable and consistent token assignments.

\subsection{Proposed Model}
\label{subsec:proposed}

While TASTE has demonstrated its ability to achieve high-quality speech reconstruction at extremely low bitrates and straightforward adaptation to joint speech–text modeling, it also suffers from a major drawback. Unlike most speech and audio tokenizers that support parallel decoding for reconstruction, TASTE relies on an AR speech decoder due to the dynamic time alignment between its tokens and speech frames. This requirement makes the decoding process rather slow, particularly for long speech signals. To address this limitation, we propose replacing the AR-based speech decoder with a discrete diffusion decoder. Interestingly, we find that this modification not only accelerates decoding but also improves reconstructed speech quality, as demonstrated in Section~\ref{sec:results}.

We follow the original structure in Fig.\ref{fig:TASTE-illustration}(a) to implement the AR-based TASTE baseline with several modifications. First, we replace the pretrained \texttt{whisper-large-v3} ASR model\cite{radford2022whisper} with \texttt{canary-180m-flash}\cite{zelasko2025training} from NVIDIA NeMo\cite{nemo_toolkit} for both encoder and aggregator initialization. Second, we use the final-layer encoder outputs for both keys and values in the aggregator’s cross-attention. Third, we proposed to use finite scalar quantization (FSQ)~\cite{mentzer2024finite} as an alternative to RVQ, which improves performance (see Section~\ref{subsec:rvq_vs_fsq}). Fourth, we adopt an encoder–decoder architecture instead of a decoder-only design. Finally, we discard the speaker embedding input, finding it unnecessary for reconstruction quality. This is consistent with findings reported in CosyVoice2~\cite{du2024cosyvoice2}.

The proposed DDM-based TASTE model shares the same architecture as the AR baseline, differing only in decoder input (Fig.~\ref{fig:TASTE-illustration}(b)): instead of conditioning on previous S3 tokens $S_{1:n-1}$, the DDM decoder takes $S_{\text{mask}}$, a partially masked version of the target sequence $S$. We follow the MDLM framework\cite{sahoo2024simple} for training but use a modified objective compared to~\eqref{eq:ddm-objective}:
\begin{equation}
    \mathcal{L}
    = \mathbb{E}_{q} \int_{t=0}^{1} 
     -\sum_{\ell} \log \big\langle 
    \mathbf{x}_\theta^{(\ell)}(\mathbf{z}_t^{1:L}, t), \mathbf{x}^{(\ell)} 
    \big\rangle \, \text{d}t,
    \label{eq:ddm-objective-biased}
\end{equation}
which is more stable and lead to consistently better results in our experiments. The overall training loss consists of the the cross entropy loss~\eqref{eq:ddm-objective-biased} and the commitment loss for the RVQ quantizer, or just the cross entropy loss~\eqref{eq:ddm-objective-biased} for the FSQ quantizer.

\begin{figure*}
    \centering    \includegraphics[width=\linewidth]{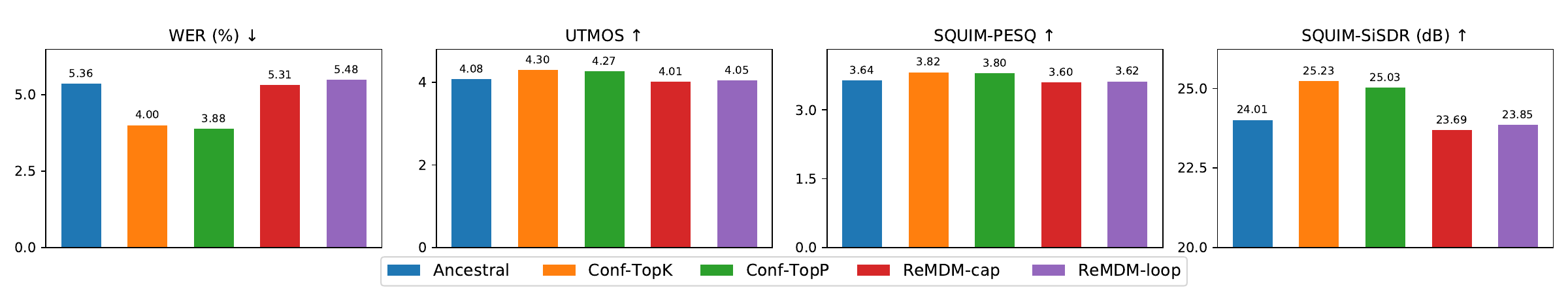}
    \vspace{-8mm}
    \caption{Comparison of different inference samplers for DDM-based speech reconstruction. Confidence-based samplers (Conf-TopK and Conf-TopP) achieve superior results compared to the Ancestral and ReMDM variants.}
    \vspace{-4mm}
    \label{fig:different-samplers}
\end{figure*}

\section{Experimental Setup}
\label{sec:experimental_setup}

\subsection{Dataset and Evaluation Metrics}
\label{subsec:dataset}

Unlike the original TASTE work trained on Emilia and LibriTTS, our models use the Granary English-only dataset~\cite{koluguri2025granary}, which contains approximately 275k hours of speech. Note that Granary is much larger but also noisier, enabling us to evaluate TASTE’s robustness on ASR-style data. For evaluation, we use the LibriSpeech \mbox{test-clean} and \mbox{test-other} sets to assess  performance under clean and noisy conditions, respectively. 

We employ a variety of metrics to evaluate speech reconstruction. Word Error Rate (WER) is computed by transcribing the reconstructed speech using NVIDIA’s FastConformer-Transducer-Large ASR model~\cite{Rekesh2023fast}. Perceptual quality is measured with \mbox{Squim-PESQ} and \mbox{Squim-SISDR}~\cite{kumar2023torchaudio}, along with \mbox{WV-MOS}\cite{Andreev2023hifi} and \mbox{UT-MOS}\cite{saeki2022utmos}. Speaker similarity is measured by extracting embeddings with WavLM\cite{chen2022wavlm} and TitaNet~\cite{Koluguri2022titanet} and computing cosine similarities, reported as \mbox{SpkSim-W} and \mbox{SpkSim-T}, respectively.

\subsection{Training and Inference Setup}
\label{subsec:training}

Training is conducted on 32 NVIDIA A100 GPUs, with each GPU processing an average batch equivalent to roughly 375 seconds of audio. Following the two-stage training procedure of the original TASTE~\cite{tseng2025taste}, we first pre-train the model without the quantizer block for 100k steps. In the second stage, we add either the FSQ or RVQ quantizer and freeze the encoder and continue training for an additional 150k steps. The total number of trainable parameters is approximately 316M. We use the Adam optimizer in both training stages, with the learning rate linearly warmed up to $5 \times 10^{-4}$ over the first 5,000 steps and subsequently decayed to $10^{-6}$ using cosine annealing. Dropout rates between 0\% and 20\% were explored, where 10\% proved most effective for AR-based TASTE, while 0\% yielded the best results for DDM-based TASTE.

For inference with DDM-based TASTE, we adopt the confidence-based Top-K (Conf-TopK) sampler from~\cite{kim2025train} as the default strategy, with the number of inference steps set to 50 unless otherwise noted. In this sampling approach, the unmasking order of tokens plays a crucial role: at each step, the model uses its own output logits to estimate confidence scores for all masked positions, then selects the $K$ tokens with the highest confidence to unmask first~\cite{kim2025train, ben2025accelerated}. The comparisons between samplers is shown in Section~\ref{subsec:sampler-choice}. We further assume that the number of S3 tokens to be predicted is known during inference. Although this information is unavailable in practice, it can be estimated using a global length predictor conditioned on the input text tokens. The effect of inaccurate length estimation is analyzed in Section~\ref{subsec:effect_of_length_estimation_error}.

\section{Results}
\label{sec:results}

In this section, we present the speech reconstruction results of our AR- and DDM-based TASTE models. We encourage readers to explore the audio samples on our demo page\footnote{\url{https://kuray107.github.io/DDMs_on_taste26_examples/demo}}to complement the quantitative results, particularly for Sections~\ref{subsec:effect_of_numer_of_steps} and~\ref{subsec:effect_of_length_estimation_error}.

\subsection{AR and DDM Baselines}
To compare AR- and DDM-based TASTE models, we evaluate both against ground-truth signals on the LibriSpeech test sets. Results are shown in Table~\ref{tab:ar-dd-comparison}. We report performance for the original speech (Original), reconstructions from ground-truth S3 sequences (S3-Oracle), models without vector quantization (-No-VQ) from the first training stage, and models with a 4-layer RVQ module (-4L-RVQ), following~\cite{tseng2025taste} with 512 tokens of 256 dimensions per layer.

As shown in Table~\ref{tab:ar-dd-comparison}, both AR- and DDM-based models perform well without vector quantization (rows 3–4), showing only slight WER increases and no loss in quality or speaker similarity compared to \mbox{S3-Oracle}. However, adding RVQ causes significant degradation for the AR-based model: WER jumps from 2.81\% to 7.60\% and UT-MOS drops from 4.15 to 3.82 on \mbox{test-clean} (row 5). In contrast, the DDM-based model shows only a modest WER rise (from 2.99\% to 5.10\%) while maintaining high UT-MOS, yielding much better reconstruction quality. Similar trends are seen on \mbox{test-other}.

Beyond reconstruction quality, DDM-based TASTE is also significantly more computationally efficient, requiring at most 50 inference steps compared to roughly 370 steps on average for AR-based TASTE. On the \mbox{test-clean} set, our DDM-based model reconstructs an utterance in about 1.65 seconds, making it \textbf{3.3$ \times$ faster} than the AR baseline, which requires an average of 5.48 seconds.

\begin{table}[t]
\centering
\setlength{\tabcolsep}{4pt}
\caption{Reconstruction performance of AR-based and DDM-based TASTE models on LibriSpeech. "No-VQ" denotes models without vector quantization, while "4L-RVQ" denotes models with a 4-layer RVQ quantizer.}
\label{tab:ar-dd-comparison}
\resizebox{\linewidth}{!}{%
    \begin{tabular}{ccccccccc}
        \toprule
        \multirow{2}{*}{\textbf{Model}} 
          & \multicolumn{3}{c}{\textbf{test-clean}} 
          & \multicolumn{3}{c}{\textbf{test-other}} \\
        \cmidrule(lr){2-4} \cmidrule(lr){5-7}
            & WER (\%) $\downarrow$ & UT-MOS $\uparrow$ &  SpkSim-W $\uparrow$
            & WER (\%) $\downarrow$  & UT-MOS $\uparrow$ &  SpkSim-W $\uparrow$\\
        \midrule
        Original     & 1.80 & 4.09 & 1.00 & 3.80 & 3.50 & 1.00 \\
        S3-Oracle    & 2.69 & 4.15 & 0.96 & 6.74 & 3.71 & 0.96 \\
        \midrule
        AR-No-VQ     & 2.81 & 4.15 & 0.96 & 8.15 & 3.73 & 0.95 \\
        DDM-No-VQ    & 2.99 & 4.23 & 0.96 & 7.50 & 3.85 & 0.96 \\
        \midrule
        AR-4L-RVQ    & 7.60 & 3.82 & \textbf{0.95} & 16.50 & 3.33 & 0.93 \\
        DDM-4L-RVQ   & \textbf{5.10} & \textbf{4.27} & 0.94 & \textbf{10.84} & \textbf{3.97} & \textbf{0.94} \\
        \bottomrule
    \end{tabular}
}
\vspace{-5mm}
\end{table}
\begin{table*}[t]
\centering
\setlength{\tabcolsep}{4pt}
\caption{Comparison of AR- and DDM-based TASTE models with different vector quantization settings (2/4/8 layers of RVQ or FSQ). DDM-based TASTE consistently outperforms AR counterparts across all metrics, and FSQ provides clear gains over RVQ.}
\label{tab:ar-dd-different-vq}
\resizebox{0.95\textwidth}{!}{%
    \begin{tabular}{cccccccccccc}
        \toprule
        \multirow{2}{*}{\textbf{AR}} 
          &  \multirow{2}{*}{\textbf{Bitrate}} 
          & \multicolumn{5}{c}{\textbf{test-clean}} 
          & \multicolumn{5}{c}{\textbf{test-other}} \\
        \cmidrule(lr){3-7} \cmidrule(lr){8-12}
          & & WER (\%) $\downarrow$ & UT-MOS $\uparrow$ & SQUIM-PESQ $\uparrow$ & SpkSim-T $\uparrow$ & SpkSim-W $\uparrow$
            & WER (\%) $\downarrow$ & UT-MOS $\uparrow$ & SQUIM-PESQ $\uparrow$ & SpkSim-T $\uparrow$ & SpkSim-W $\uparrow$ \\
        \midrule
        2L-RVQ       & 95   & 10.07 & 3.70 & 3.53 & 0.64 & 0.93 & 16.22 & 3.20 & 3.23 & 0.60 & 0.92 \\
        2L-FSQ       & 95   & 6.36  & 3.93 & 3.55 & 0.62 & 0.93 & 12.20 & 3.50 & 3.25 & 0.58 & 0.91 \\
        4L-RVQ       & 190  & 7.60  & 3.82 & 3.56 & 0.68 & \textbf{0.95} & 16.50 & 3.33 & 3.23 & 0.64 & 0.93 \\
        4L-FSQ       & 190  & \textbf{4.87}  & 4.07 & 3.65 & 0.66 & 0.94 & \textbf{10.70} & 3.67 & 3.37 & 0.63 & 0.93 \\
        8L-RVQ       & 380  & 7.94  & 3.92 & 3.59 & \textbf{0.71} & \textbf{0.95} & 16.47 & 3.39 & 3.25 & 0.67 & 0.94 \\
        8L-FSQ       & 380  & 5.14  & \textbf{4.13} & \textbf{3.70} & \textbf{0.71} & \textbf{0.95} & 10.94 & \textbf{3.71} & \textbf{3.40} & \textbf{0.68} & \textbf{0.95} \\
        \midrule \midrule
        \multirow{2}{*}{\textbf{DDM}} 
          &  \multirow{2}{*}{\textbf{Bitrate}} 
          & \multicolumn{5}{c}{\textbf{test-clean}} 
          & \multicolumn{5}{c}{\textbf{test-other}} \\
        \cmidrule(lr){3-7} \cmidrule(lr){8-12}
          & & WER (\%) $\downarrow$ & UT-MOS $\uparrow$ & SQUIM-PESQ $\uparrow$ & SpkSim-T $\uparrow$ & SpkSim-W $\uparrow$
            & WER (\%) $\downarrow$ & UT-MOS $\uparrow$ & SQUIM-PESQ $\uparrow$ & SpkSim-T $\uparrow$ & SpkSim-W $\uparrow$ \\
        \midrule
        2L-RVQ   & 95  &  5.18    &  4.32     &   \textbf{3.84}    & 0.63     &  0.93    &  8.50    &  \textbf{4.08}     &   3.65    &  0.59    &  0.91    \\
        2L-FSQ   & 95  &  \textbf{3.95}    &  \textbf{4.31}     &   \textbf{3.84}    & 0.62     &  0.92    &  \textbf{7.51}    &  \textbf{4.08}     &   \textbf{3.66}    &  0.56    &  0.90    \\
        4L-RVQ   & 190 &  5.10    &  4.27     &   3.80    & 0.68     &  0.94    &  10.84   &  3.97     &   3.58    &  0.65    &  0.94    \\
        4L-FSQ   & 190 &  4.00    &  4.30     &   3.82    & 0.67     &  \textbf{0.95}    &  8.62    &  4.00     &   3.60    &  0.64    &  0.94    \\
        8L-RVQ   & 380 &  4.20    &  4.24     &   3.78    & \textbf{0.71}     &  \textbf{0.95}    &  9.39    &  3.89     &   3.54    &  \textbf{0.69}    &  0.94    \\
        8L-FSQ   & 380 &  3.99    &  4.14     &   3.79    & \textbf{0.71}     &  \textbf{0.95}    &  9.08    &  3.90     &   3.54    &  \textbf{0.69}    &  \textbf{0.95}    \\
        \bottomrule
    \end{tabular}
}
\vspace{-6mm}
\end{table*}
\subsection{Effect of Vector Quantization}
\label{subsec:rvq_vs_fsq}
To study the impact of vector quantization, we trained AR and DDM variants with different VQ configurations, comparing RVQ and FSQ modules with 2, 4, or 8 layers, yielding six variants per model. Results in Table~\ref{tab:ar-dd-different-vq} show that across all configurations, DDM-based TASTE consistently outperforms its AR counterparts, particularly in WER, demonstrating superior efficiency and reconstruction quality. FSQ also clearly outperforms RVQ, especially for AR models, reducing WER by about 35\% on \mbox{test-clean} and 31\% on \mbox{test-other}, while improving UT-MOS by +0.14 and +0.31, respectively. For DDM, RVQ already performs strongly, so FSQ provides smaller but consistent WER gains. Lastly, while using fewer VQ layers degrades AR performance across all metrics, DDM remains robust, though speaker similarity drops noticeably.

\subsection{Effect of Sampler Choice}
\label{subsec:sampler-choice}

Next, to evaluate the impact of sampler choice on DDM inference, we compare the default confidence-based TopK (Conf-TopK) sampler with four alternatives: Ancestral~\cite{sahoo2024simple}, confidence-based TopP (Conf-TopP)\cite{yu2025dimple}, and two ReMDM variants (ReMDM-Cap and ReMDM-Loop)\cite{wang2025remasking}, which allow non-masked tokens to be remasked during the denoising process. Results with the DDM-4L-FSQ TASTE model are shown in Fig.~\ref{fig:different-samplers}, where confidence-based samplers (Conf-TopK and Conf-TopP) achieve the best performance, while Ancestral and ReMDM perform worse—despite often being reported to produce more diverse outputs in unconditional text generation\cite{sahoo2024simple, wang2025remasking}. This is likely because confidence-based greedy sampling introduces a strong bias toward highly probable tokens, which can hurt tasks like unconditional text generation where diversity is crucial and context is limited. In our setting, however, speech reconstruction is strongly conditioned on text and TASTE embeddings, making diversity less important. This strong conditioning makes confidence-based sampling more effective, consistently producing higher-quality reconstructions than more stochastic methods.

\begin{table}[t]
    \caption{Reconstruction performance of DDM with different numbers of inference steps on test-clean.}
    \vspace{-3mm}
    \label{tab:ddm-steps}
    \begin{center}
        \setlength\tabcolsep{4.0pt}
        \resizebox{\columnwidth}{!}{%
            \begin{tabular}{cccccc}
                \toprule
                \textbf{\# Steps} & \textbf{Inference Time (s)} & \textbf{WER (\%)} $\downarrow$ & \textbf{UTMOS} $\uparrow$ & \textbf{SQUIM-PESQ} $\uparrow$ & \textbf{SpkSim-W} $\uparrow$ \\
                \midrule
                1   & 1.09 & 5.14 & 3.81 & 3.39 & 0.95 \\
                \midrule
                10  & 1.18 & 3.70 & 4.28 & 3.80 & 0.95 \\
                25  & 1.36 & 3.83 & 4.29 & 3.81 & 0.95 \\
                50  & 1.65 & 4.00 & 4.30 & 3.82 & 0.95 \\
                100 & 2.29 & 4.01 & 4.30 & 3.82 & 0.94 \\
                \bottomrule
            \end{tabular}
        }
    \end{center}
    \vspace{-10mm}
\end{table}

\subsection{Effect of Number of Inference Steps}
\label{subsec:effect_of_numer_of_steps}
As discussed in Section~\ref{subsec:ddm-intro}, a major advantage of DDMs over AR models is the flexibility to adjust the number of inference steps, trading off efficiency and generation quality. To evaluate this, we tested the DDM-4L-FSQ model with 1, 10, 25, 50, and 100 steps. The Conf-TopK sampler is used in all cases except for single-step inference, where the sampler choice has no effect. As shown in Table~\ref{tab:ddm-steps}, the model achieves strong performance with as few as 10 steps, showing no degradation across metrics. In fact, 10 steps yield the lowest WER, underscoring DDM’s ability to provide both efficiency and high-quality reconstruction. Even single-step inference remains usable, with only modest degradations in WER (from 4.00\% to 5.14\%) and UT-MOS (from 4.30 to 3.81), demonstrating that DDMs can deliver significantly faster speech reconstruction than AR baselines while maintaining high quality.

\subsection{Effect of Length Estimation Errors}
\label{subsec:effect_of_length_estimation_error}
\begin{figure}
    \centering
    \includegraphics[width=\linewidth]{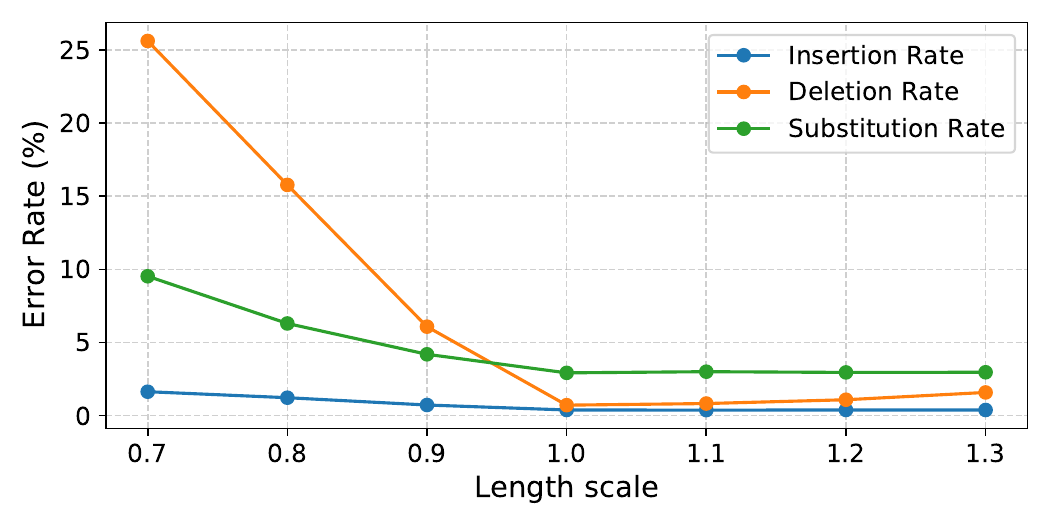}
    \vspace{-4mm}
    \caption{Effect of length scale mismatch on insertion, deletion, and substitution rates.}
    \vspace{-5mm}
    \label{fig:length_scale}
\end{figure}

Finally, we revisit the assumption in Section~\ref{subsec:training} that the model has access to the global S3 token length during inference, which in practice requires an auxiliary length predictor. To analyze its impact, we evaluate the DDM-4L-FSQ model under length scales ranging from 70\% to 130\% of the original sequence, measuring insertion, deletion, and substitution rates. Results are shown in Fig.~\ref{fig:length_scale}. When the inference length is shorter than the ground truth, the model maintains the same speaking rate and truncates the sequence tail, causing higher deletion and substitution rates, while insertion remains stable. In contrast, when the length is longer, it simply appends silence tokens with minimal ASR degradation. These results suggest that overestimation is relatively safe, whereas underestimation leads to severe errors. A promising future direction is to extend the DDM to predict an end-of-sequence (EOS) token\cite{nie2025llada}, eliminating the need for an external length predictor.

\section{Conclusion}
\label{sec:conclusion}
In this paper, we presented the first comprehensive study of applying discrete diffusion to speech tokenization and reconstruction within the TASTE framework. The proposed discrete diffusion-based decoder achieves both higher reconstruction quality and substantially faster inference than the AR baseline, requiring only ten denoising steps without performance degradation and even supporting single-step generation with minor degradation. Experiments further demonstrated FSQ outperforms RVQ in quantizer design, confidence-based samplers are better suited for speech reconstruction than more stochastic alternatives, and DDMs remain robust under varying inference lengths. These results establish DDM-based TASTE as a more efficient and flexible alternative to the AR baseline, offering practical benefits for future speech applications.

\newpage
\footnotesize
\bibliographystyle{IEEEbib}
\bibliography{strings,refs}

\end{document}